\documentstyle[12pt]{article}
\newcommand{\EQ}{\begin{equation}}
\newcommand{\EN}{\end{equation}}
\newcommand{\EQN}{\begin{eqnarray}}
\newcommand{\ENN}{\end{eqnarray}}
\newcommand{\CR}{\nonumber \\}

\newcommand{\A}{\alpha}

\newcommand{\lm}{\lambda}

\newcommand{\T}{\theta}
\newcommand{\E}{\epsilon}

\newcommand{\SL}{\ ^{*} {\cal L}}

\newcommand{\RE}{{\cal R}}
\newcommand{\SR}{\ ^{*} {\cal R}}
\newcommand{\NA}{{\cal N}}
\newcommand{\SN}{\ ^{*} {\cal N}}
\newcommand{\SM}{\ ^{*} {\cal M}}
\newcommand{\MON}{{\rm Mon}(r|\SL)}
\newcommand{\MONP}{{\rm Mon}(r'|\SL)}
\newcommand{\SE}{\ ^{*}\varepsilon}
\newcommand{\DTX}{ d_\T^x(\vec r)}
\newcommand{\DTY}{ d_\T^y(\vec r)}
\newcommand{\SDX}{\ ^{*} d_\T^x(\vec r)}
\newcommand{\SDY}{\ ^{*} d_\T^y(\vec r)}

\begin{document}
\newfont{\elevenmib}{cmmib10 scaled\magstephalf}
\renewcommand{\thefootnote}{\fnsymbol{footnote}}

\renewcommand{\thefootnote}{\arabic{footnote}}
\setcounter{footnote}{0}
\baselineskip=0.8cm

\begin{center}
\vskip20pt
\large

{\bf Physical Equivalence on Non-Standard Spaces

and 

Symmetries on Infinitesimal-Lattice Spaces}

\end{center}

\begin{center}
\vskip20pt

Tsunehiro KOBAYASHI
\vskip3pt  
{\it  Department of General Education, Tsukuba College of Technology, 

Tsukuba, 305-0005 Ibaraki, Japan}

E-mail: kobayash@a.tsukuba-tech.ac.jp
\vskip10pt

\end{center}
\begin{abstract}
{\small 
Equivalence in physics is discussed on the basis of experimental data 
accompanied by experimental errors. 
The introduction of the equivalence being consistent 
with  the mathematical definition 
is possible only in theories constructed 
on non-standard number spaces 
by taking the experimental errors 
as infinitesimal numbers of the non-standard spaces. 
Following the idea for the equivalence (the physical equivalence), 
a new description of space-time in terms of infinitesimal-lattice points 
on non-standard real number space $\SR$ is proposed. 
The infinitesimal-lattice space, $^*{\cal L}$, is  represented by the set of 
points on $\SR$ which are written by $l_n=n\SE$, where the infinitesimal 
lattice-spacing $\SE$ is determined by a non-standard natural number $^*N$ 
such that $\SE\equiv \ ^*N^{-1}$. 
By using infinitesimal neighborhoos ($\MON$) of real number $r$ on $\SL$ 
we can make a space $\SM$ which is isomorphic to $\RE$ as additive group. 
Therefore, every point on $(\SM)^N$ automatically has 
the internal confined-subspace $\MON$. 
A field theory on $\SL$ is proposed. 
To determine a projection from $\SL$ to $\SM$, a fundamental principle based 
on the physical equivalence is introduced. 
The physical equivalence is expressed by the totally 
equal treatment for indistinguishable quantities in 
our observations. 
Following the principle, we show that $U(1)$ and $SU(N)$ symmetries 
on the space $(\SM)^N$ are induced from the internal substructure 
$(\MON)^N$. 
Quantized state describing configuration space is constructed on $(\SM)^N$. 
By providing that the subspace $(\MON)^N$ is local inertial system 
of general relativity, infinitesimal distance operators are consistently 
introduced. 
We see that 
Lorentz and general relativistic transformations are also represented by 
operators which involve the $U(1)$ and $SU(N)$ internal symmetries. 
}

\end{abstract}

\vskip100pt
(To appear in Proceedings ``Symmetries in Science XI`` (Plenum, New york).
2000)

\hfil\break
{\large {\bf 1. Introduction}

--Why are non-standard spaces needed in 
theories of physics?--}
\vskip10pt

For our recognition derived from observations 
the judgment of equivalence between two or more phenomena 
plays a very important role. 
It is kown that the equivalence is rigorously defined 
in mathematics in terms of the following 
three conditions; 

(1) $A \sim A\, $ (reflection)

(2) $A \sim B \Longrightarrow B \sim A,$ (symmetry)

(3) $A \sim B,\ \ B \sim C \Longrightarrow A \sim C. $ (transitivity)
\hfil\break
In observations of physics, that is, in experiments, 
the equivalence (physical equivalence) can be described as follows:
\begin{center}
{\it Two phenomena {\rm A} and {\rm B} are equivalent, 

if {\rm A} and {\rm B} coincide within the experimental errors.}
\end{center}
It should be stressed that the physical equivalence is detemined by the experimental errors. 
Futhermore we must recognize that there is no experiments accompanied by 
no error. 
We should consider that experimental errors are one of the fundamental 
observables in our experiments. 
It is quite hard to understand that there is no theory which involves 
any description of experimental errors, even though they 
are very fundamental observables. 
It is also hard to understand that 
the question whether such physical equivalence is compatible 
with the mathematical definition represented by the above three 
conditions had never been discussed.  
Let us discuss the question here. 
We easily see that the first two conditions, that is, reflection and symmetry 
are compatible with the physical equivalence based on experimental errors. 
We can, however, easily 
present examples which break the third condition (transitivity), 
that is to say, 
$A\sim B$ and $B\sim C$ are satisfied within their errors 
but $A$ and $C$ does not coincide within their errors. 
This arises from the fact that real numbers which exceed any real numbers 
can be made from repeated additions of a non-zero real number 
because of Archimedian property of real number space. 

{\it How can we introduce the physical equivalence in theories?} 
\hfil\break
Consistent definition of the physical equivalence is allowed, only when 
experimental errors are taken as {\it infinitesimal numbers}[1] 
in non-standard spaces. 
This result comes from the fact that any non-zero real numbers cannot be made 
from any finite sum of infinitesimal numbers. 
Any repetitions of the transitivity, that is, repeated additions of 
any infinitesimal numbers does not lead any non-zero 
real numbers. 
We can describe the situation as follows;
$$
\forall\E \in {\rm Mon}(0)\ \ {\rm and}\ \ \forall N \in {\cal N} 
\Longrightarrow \E N \in {\rm Mon}(0),
$$
where Mon(0) and ${\cal N}$, respectively, stand for the set of all 
infinitesimal numbers on non-standard spaces and the set of all natural 
numbers. 
From the above argument we can conclude that 
we must make theories, in which the physical equivalence based on 
experimental errors is described in terms of 
the mathematically consistent form, on a non-standard space. 
This is the reason why non-standard spaces are needed in the description of 
realistic theories based on the physical equivalence. 
It is once more stressed that such realistic theories 
must involve the fundamental observables, experimental 
errors, in the mathematically rigorous way. 

An example for the introduction of the physical equivalence in quantum 
mechanics on non-standard space has been presented in the derivation 
of decoherence between quantum states for the description of quantum theory 
of measurements.[2-4] 
In the theory not only the decoherence required for the wave function collapse 
but also that for microcanonical ensembles of statistical mechanics 
(principle of {\it a priori} equal probabilities) have been simultaneously 
derived by the realization of the physical equivalence. 
Though we have many other interesting problems for the construction of 
theories 
on non-standard spaces,[5-12] we shall investigate space-time structure and 
field theory,[11,12] following the idea of the physical equivalence 
based on experimental errors, in this paper.
\vskip5pt

To help to see contents of this paper, we present a list of sections here: 
\hfil\break
{\bf 1. Introduction 
\hfil\break
2. On observation of continuity of space-time
\hfil\break
3. Short review of some fundamentals of non-standard space
\hfil\break
4. Infinitesimal-lattice spaces $\SL$
\hfil\break
5. Translarions and rotations on $\SM$
\hfil\break
6. Confined fractal-like property of $\SL$
\hfil\break
7. Construction of fields on $\SM$ 
\hfil\break
8. Internal symmetries on $(\SM)^N$ induced from the confined 
substructure 

$(\MON)^N$ 
\hfil\break
9. Quantized configuration space and infinitesimal distances 
\hfil\break
10. Translations, rotations and Lorentz and general relativistic transformations 
\hfil\break
11. Remarks on fermionic oscillators 
\hfil\break
12. Concluding remarks }

\vskip10pt
\hfil\break
{\large {\bf 2. On observation of continuity of space-time}}
\vskip10pt

Space-time structure has been studied as one of exciting theme in physics. 
Whether space-time is continuous (as represented by the set of 
real numbers $\RE$) or discrete (as represented by the set of discrete 
lattice-points) 
is a fundamental question for the space-time structure. 
We may ask 

``{\it How can we experimentally verify the continuous property of space-time?''. }
\hfil\break
As noted in the first section, 
we have no experiment accompanied by no error. 
Taking into account that 
experimental errors are fundamental observables in physical phenomena, 
we should understand that the continuity of space-time cannot be directly 
verified in any experiments. 
This means that a discrete space-time is sufficient to describe realistic 
space-time. 
We, however, know that translational and rotational invariances (including 
Lorenz invariance) with respect to
space-time axes seems to be very fundamental 
concepts in nature and lattice spaces break them. 
This disadvantage seem to be very difficult to 
overcome on usual lattice spaces having a finite 
lattice-spacing between two neighboring lattice-points. 
As was discussed in the introduction, 
experimental errors must be described in terms of infinitesimal numbers on 
non-standard spaces. 
On non-standard spaces[1] we can introduce infinitesimal lengths 
which are smaller than all real numbers except $0$. 
It will be an interesting question whether we can overcome the disadvantage on 
lattice spaces defined by infinitesimal lattice-spacing. 
Actually such infinitesimal discreteness cannot be observed in our experiments, 
where all results must be described by real numbers. 
This fact indicates that such lattice space-time will possibly be observed 
as continuous structure. 
Hereafter we call lattice spaces discretized by infinitesimal numbers 
{\it infinitesimal-lattice spaces} and they are denoted by $\SL$.[11] 
That is to say, 
such a lattice space $\SL$ is constructed as the set of non-standard numbers 
which are separated by an infinitesimal lattice-spacing $\SE$
on $\SR$(the non-standard extension  of $\RE$). 
Lattice-points on $\SL$ are defined by 
$$
l_n=n \SE,\ \ \ \ {\rm for}\ \ n\in\ \SN
$$ 
where $\SN$ stands for the non-standard extension of the set of natural 
numbers $\NA$(=0, 1, 2, 3, $\cdots$) and consists of all natural numbers 
and non-standard natural numbers which are infinity. 
It is transparent that such $\SL$ do not contain many of real numbers. 
There is, however, a possibility that parts of infinitesimal neighborhoods of 
all real numbers are contained in $\SL$, because it is known that 
the power of $\SL$ is  same as that of $\RE$.[1] 
If it is true, there is a posibility 
that a space constructed from the set of all 
infinitesimal neighborhoods on $\SL$ will be isomorphic to $\RE$ 
and translations and rotations on the space can be introduced as same as 
those on $\RE$.[11] 
In this paper we shall start from the investigation of 
properties of $\SL$ and examine the construction of 
a new theory on the space-time represented by $\SL$. 
If we can succeed it, we shall construct a field theory on the new space.[12] 

\vskip10pt
\hfil\break
{\large {\bf 3. Short review of some fundamentals of non-standard space}}
\vskip10pt

Here we shall briefly review some fundamental languages of non-standard 
analysis,[1] which are not familiar to physicists but needed in the 
argument of this paper. 
Readers who are familiar to non-standard analysis may skip this section 
and go to the next section. 

\hfil\break
{\bf (i) Free ultra-filters}
\hfil\break
The set of real numbers $\cal R$ can be extended to the set of numbers 
($^*\cal R$) containing infinitesimal and infinity   in terms of 
free ultra-filters($\cal F)$ over $\cal N$
(${\cal N}=(0,1,2,....)$ denotes the set of natural numbers). 
The free ultra-filters satisfy the following properties: \hfil\break
(a) $ {\cal N} \in \cal F,\ \ \ \phi$(empty set)$\not\in \cal F$, \hfil\break
(b) $A,B\in {\cal F}\ \Longrightarrow \ A\cap B \in \cal F$, \hfil\break
(c) $A\in{\cal F},\ \ A\subseteq B\ \Longrightarrow \ B\in \cal F$, \hfil\break
(d) $\cal F$ contains no finite set, (the filter having 
this property is called free),\hfil\break
(e) either $E\in \cal F$ or ${\cal N}-E\in\cal F$ for $^\forall E\subseteq 
\cal N$ (the filter having 
this property is called ultra-filter over $\cal N$).\hfil\break
Hereafter filters and ultra-filters always mean free ultra-filters. 

\hfil\break
{\bf (ii) Equivalence in terms of free ultra-filter and non-standard extension}
\hfil\break
We can construct the non-standard extension of $\cal R$ by introducing an 
equivalence relation on sequences in ${\cal R}^{\cal N}$ by means of 
an  ultra-filter $\cal F$. 
The equivalence relation, $\sim_{\cal F}$, is defined as follows;
\EQ
  f\sim_{\cal F} g
\EN 
if and only if 
$\{n\in{\cal N}|f(n)=g(n)\}\in \cal F$, 
where $f$ and $g$ are, respectively, represented by ultra-product 
\EQ
  f=\prod_{n\in {\cal N}} f(n),\ \ \ \ \ g=\prod_{n\in {\cal N}} g(n). 
\EN 
Note that the sequences associated with the equivalence relation may be 
expressed by using ultra-powers 
\EQ
\prod_{n\in\ \cal N}f(n)/\sim_{\cal F}.
\EN 
We may write the non-standard 
extension of $\cal R$ in terms of the quotient space 
\EQ
  ^*{\cal R}=\cal R^{\cal N}/\sim_{\cal F}.
\EN 
We also have non-standard extensions of $\cal N$, 
$\cal Z$(the set of integers), 
$\cal Q$(the set of rational numbers),
$\cal C$( the set of complex 
number) and so on, which are denoted as $^*\cal N$, $^*\cal Z$,
$^*\cal Q$, $^*\cal C$ and so forth, respectively. 
It is shown that ${\cal R}\subset \ ^*\cal R$ 
and the magnitudes of the non-standard natural numbers, 
$^{\forall *}N \in\ ^*{\cal N}- \cal N$, are infinity.  

\hfil\break
{\bf (iii) Definitions of $\leq,\  +$ and $\times$}
\hfil\break
We can introduce the order $\leq$ between two ultra-products $f$ and $g$
as follows;

if and only if $\{n\in{\cal N}|f(n)\leq g(n)\}\in \cal F$, 
\EQ
f\leq g.
\EN
It is shown that $\SN$, $^*{\cal Z}$, $^*{\cal Q}$
and $\SR$ are totally ordered sets. 
\hfil\break
Sum and multiplication are defined by 
\EQ
f+g=\prod_{n\in \NA}(f(n)+g(n)), \ \ \ f\times g=\prod_{n\in \NA}(f(n)\times 
g(n)).
\EN 
An example of $^*N \in \SN$ is given by using the ultra-product 
\EQ
^*N=\prod_{n\in \NA} (n+1).
\EN 
Following the order $\leq$ defined by free ultra-filters $\cal F$ and 
the properties (d) and (e) of $\cal F$, 
it is obvious that 
\EQ
^*N>N,\ \ \  {\rm for }\ \forall N\in \NA
\EN 
because the set $\{ n\in \NA|n+1\leq N\}\not\in {\cal F}$ is 
a finite set on $\NA$, whereas that of 
$\{n\in \NA|n+1>N\}\in {\cal F}$ is an infinite set on $\NA$. 

\hfil\break
{\bf (iv) Standard part map(st-map)}
\hfil\break
We have a projection of every finite number($^*r$) of $^*\cal R$ 
to a unique element($r$) 
of $\cal R$, which is called as the standard part map(st-map) 
and written by 
\EQ
   {\rm st}(^*r)=r.
\EN 
All infinitesimal numbers are mapped at zero.  
\vskip1pt
\hfil\break
{\bf (v) Monad of $r\in\cal R$ (Mon($r$))}
\hfil\break
Each real number $r\in \cal R$ has its own infinitesimal neighborhood 
Mon($r$) which is called monad of $r$ and defined 
by the set of $^*r\in\ ^*\cal R$ satisfying 
\EQ 
{\rm st}(^*r-r)=0. 
\EN 
In other words it may be represented by the set of $^*r\in\ ^*\cal R$ 
satisfying st$(^*r)=r$. We see that Mon($0$) contains all infinitesimals.  
Note that the map of all elements being finite into monads of real numbers 
is unique, that is, no element of $\SR$ cannot belong two or more monads 
simultaneously such that 
$$
{\rm Mon}(r)\cap {\rm Mon}(r')=\phi, \ \ \ {\rm for}\ \ r\not=r'
\ and r,r'\in {\cal R}. 
$$
\vskip1pt
\hfil\break
{\bf (vi) Powers of $\SN$ and $\SR$}
\hfil\break
Powers of $\SN$ and $\SR$ are same as that of $\RE$, that is, 
$\SN$ and $\SR$ have the same continuous power as that of $\RE$. 
In fact $\SR_\RE/{\rm Mon}(0)\cong \RE$ as field is shown,[1] 
where $\SR_\RE$ is the set of elements of $\SR$, of which elements 
$^*r$ are 
satisfied by the relation st$(^*r)\in \RE$. 

\vskip10pt
\hfil\break
{\large {\bf 4. Infinitesimal-lattice spaces $\SL$}}
\vskip10pt

Let us take a non-standard natural number 
\EQ
^*N\in\ \SN-\NA,
\EN 
which is an infinity.[1] 
We take the closed set $[-\ ^*N/2,\ ^*N/2]$ on $^*\RE$ and put $(\ ^*N)^2
\ +\ 1$
points with an equal spacing $\SE=\ ^*N^{-1}$ on the set. 
For the convenience of the following discussions $\ ^*N$ is chosen as 
$\ ^*N/2 \in ^*{\cal N}$. 
The length between two neighboring points is $\SE$ which is 
an infinitesimal, i.e. $\ ^*\E \in$Mon(0). 
Let us consider the set of the infinitesimal lattice-points $\SL$,[11] 
which consists 
of these $(\ ^*N)^2\ +\ 1$ discrete points on the closed set. 
Lattice-points on $\SL$ are written by 
\EQ
l_n=n  \SE, 
\EN 
where $n\in \ ^*{\cal Z}$ 
and fulfil the relation 
\EQ
-(\ ^*N)^2/2\leq n \leq (\ ^*N)^2/2.
\EN 
We put two end points as the same point, i.e.
\EQ
l_{(\ ^*N)^2/2}=\ ^*N \ = l_{-(\ ^*N)^2/2}=-\ ^*N .
\EN 
This choice corresponds to the choice of periodic boundary which is required 
for the introduction of translations on $\SL$. 
We may consider $\SL$ as the set of
  $(\ ^*N)^2$ points with the equal spacing $\SE$ 
on the circle of the radius $\ ^*N/2\pi$. 
From the process of the construction of $\SL$ it is transparent that 
\EQ
\SL\not\supset \RE.
\EN 
Actually it is obvious that all irrational numbers of $\RE$ are not contained 
in $\SL$, because $\ ^*N$ is taken as an element of $\SN$ and 
$n\ ^*\E=n/\ ^*N$ is an element of $\ ^*{\cal Z}$. 

Let us show a theorem:\hfil\break
{\it Monads of all real numbers, {\rm Mon}($r$) $\forall r \in \RE$, 
have their elements on $\SL$. }

Proof: Take a real number $r\in \RE$. 
The number $r$ is contained in the closed set $[-\ ^*N/2,\ 
^*N/2]$ on $\SR$, because $^*N$ is an infinity of $\SN$ and then 
$[-\ ^*N/2,\ ^*N/2]\supset \RE$. 
Since the lattice-points of $\SL$ divide the closed set
into $( ^*N)^2$ regions of which 
lenght is $\SE$, the real number $r$ must be on a lattice-point or between 
two neighboring lattice-points whose distance is $\SE$. 
We can, therefore, find out a non-standard integer $N_r$ fulfilling the 
following relation;
\EQ
N_r \SE\leq r< (N_r+1) \SE,
\EN 
where $|N_r|\in\SN-\NA$. 
The difference $r-N_r \SE$ is an infinitesimal number smaller than $\SE$. 
Thus we can define the infinitesimal neighborhood of $r$ on $\SL$ such that 
\EQ
{\rm Mon}(r|\SL)\equiv \{ l_n(r)=(N_r+n)\SE|n\in \ ^*{\cal Z},\ n\SE\in {\rm Mon}(0)\}. 
\EN
The relation 
\EQ
{\rm st}(l_n(r))=r
\EN 
is obvious. 
The theorem has been proved. 
Hereafter we shall call Mon($r|\SL$) and its elements $l_n(r)$ 
monad lattice-space ($\SL$-monad) and monad lattice-points, respectively. 

From the above argument we see that there is one-to-one correspondence between 
$\RE$ and 
$$ 
\SL_{l(\RE)}\equiv \{ l_0(r)|r\in \RE\}
$$ 
(the set of $l_0(r)$ 
for $\forall r\in \RE$) with respect to the correspondence $r\leftrightarrow
l_0(r)$. 
Note also that from the definition of monad we have the relations 
\EQ 
\MON \cap \MONP=\phi,\ \ \ {\rm for}\ \ r\not=r',\ r,r'\in \RE.
\EN 

Magnitudes of lattice-points contained in all of the monad lattice-space 
Mon($r|\SL$) for $\forall r\in \RE$ are not infinity,  
because they are elements of monads of real numbers. 
We shall write the set of all these finite lattice-pionts by 
$$
\SL_\RE\equiv \{ l_n(r)|r\in\RE,\ n\in\ ^*{\cal Z},\ n\SE\in {\rm Mon}(0)\}= 
\cup_{r\in \RE}\ \MON. 
$$
The sets $\SL_\RE$ and Mon$(0|\SL)$ are additive groups. 
Note here that $\SL_{l(\RE)}$ is not an additive group, 
because in general $l_0(r)+l_0(r')\not=l_0(r+r')$ possibly happens, 
that is, $N_{r+r'}$ is not always equal to $N_r+N_{r'}$
but possibly equal to $N_r+N_{r'}+1$.(See the 
definition of $N_r$ given in (16).) 
It is apparent that 
\EQ
  \SL_\RE=\SL_{l(\RE)}+{\rm Mon}(0|\SL)\ \ {\rm and}\ \  
\SL_{l(\RE)}\cap {\rm Mon}(0|\SL)=\{ 0\}. 
\EN  
Let us introduce the quotient set of $\SL_\RE$ by ${\rm Mon}(0|\SL)$
as 
$$
\SM\equiv \SL_\RE/{\rm Mon}(0|\SL).
$$ 
From one-to-one correspondence between $\RE$ and $\SL_{l(\RE)}$ and 
the relations (20) we see that there is one-to-one correspondence 
between $\RE$ and $\SM$, and thus 
\EQ
\SM \cong \RE
\EN 
as additive groups, where the addition on $\SM$ may be described
by st-map of the addition  on $\SL_\RE$ such that 
$ {\rm st}(l_n(r)+l_m(r'))=r+r'$ 
for $\forall l_n(r)\in\MON$ and $\forall l_m(r')\in \MONP$ with $r,r'\in \RE$.

We can construct the same quotient set where the zero point  
of the subset $\SL_\RE$ is taken at an arbitrary point of $\SL$. 
That is to say, by using the relative distance $l_{Nm}$ 
between an arbitrary point $l_m$ and the origin $l_N$ as 
\EQ
l_{Nm}\equiv l_m-l_N=(m-N)\SE, 
\EN 
we can proceed the same argument for constructing $\SM$. 
This means that $\SL$ contains infinite number of subsets
which are congruent to $\SM$. 
When we consider $\SL$ on the circle with the radius $\ ^*N/2\pi$ on $\SR^2$, 
the angle of the sector including one $\SM$ is infinitesimal. 
This means that $\SM$ can be taken as a straight line on two dimensional
real space $\RE^2$,  even if it is put on the circle of $\SR^2$. 

Finally we summarize the notations newly introduced in this section 
for the convienience in the following discussions:

$\SL=$the set of all infinitesimal lattice-points, 
(infinitesimal latice-space) 

Mon($r|\SL$)=the set of lattice points which are elements of Mon($r$) 
for $r\in \RE$, 

(monad lattice-space) 

$\SL_\RE=$the set of lattice-points which are elements of Mon$(r|\SL)$
for $\forall r\in \RE$, 

(finite infinitesimal lattice-space) 

$\SM=\SL_\RE/{\rm Mon}(0|\SL)$, (observed space). 

\vskip10pt
\hfil\break
{\large {\bf 5. Translations and rotations on $\SM$}}

\vskip10pt

Since $\SM\cong \RE$ as additive groups has been proved in the last section, 
it is obvious that translations and rotations on $\SM$ can be taken 
as same as those on $\RE$. 
We shall here study translations and rotations on the sub-lattice space
$\SL_\RE$ and construct them on $\SM$ explicitly.(See Ref.11.) 

\hfil\break
{\bf 5.1 Translations }

In general a translation $^*\hat p_m$ on $\SL$ is represented by 
the following map from $\SL$ to $\SL$; 
\EQ
^*\hat p_m l_n=l_{n+m},\ \ \  {\rm for}\ \ n,m \in \ ^*{\cal Z}.
\EN 
The displacement length by this translation is 
\EQ
d_m\equiv l_{n+m}-l_n=l_m=m \SE.
\EN 
Let us study only finite translations restricted by 
\EQ
{\rm st}(d_m)\in \RE.
\EN 
Note that the subset of $\SL$, i.e., $\SL_\RE$, is mapped on to 
$\SL_\RE$ by these finite translations. 
We may, therefore,
consider that these finite translations represent translations on $\SL_\RE$. 
We also see that under these translations all the elements of 
$\MON$ are replaced on those of $\MONP$, where
\EQ
r'={\rm st}(r+d_m).
\EN 
Let us show that the elements of $\MON$ and those of $\MONP$ 
have one-to-one correspondence. 
The displacement of $\MON$ to $\MONP$ for $r$, $r'\ \in \RE$ is 
described by 
\EQ
d_{rr'}=(N_{r'}-N_r) \SE,
\EN 
where 
$N_r \SE\leq r< (N_r+1) \SE$ and  $N_{r'} \SE\leq r'< (N_{r'}+1) \SE$ 
with $N_r,\ N_{r'}\ \in \ ^*{\cal Z}$. 
It is trivial that st$(d_{rr'})=r'-r$  is finite. 
An element of $\MON$, $l_m(r)=(N_r+m)\SE$, is replaced on an element of 
  $\MONP,\ l_m(r')=(N_{r'}+m) \SE$, by the translation. 
Considering the inverse of the translation, which is
described by the displacement  $-d_{rr'}$,
one-to-one correspondence between $\MON$ and $\MONP$ is obvious. 

We can change $d_{rr'}$ by infinitesimal length $\Delta d_k=k\SE$ such that 
$d_{rr',k}\equiv d_{rr'}+\Delta d_k$, where $k$ must be taken as integers 
of $ \ ^*{\cal Z}$ satisfying the relation $k\SE\in {\rm Mon}(0)$. 
Note that $\Delta d_k$ does not depend on $r$ and $r'$. 
One-to-one correspondence is not affected by these infinitesimal 
changes. 
This fact means that all infinitesimal translations($^*\hat p_0(\SL_\RE)$) 
described by $\Delta d_k$ are 
mapped on the zero translation on $\SM$. 
Thus we see that all the translations from $\MON$ to $\MONP$ on $\SL_\RE$, 
which produce the displacement st$(d_{rr'})$ on $\SM$, 
  are expressed by 
\EQ
^*\hat p_m(\SL_\RE)\equiv\ ^*\hat p_m+\ ^*\hat p_0(\SL_\RE),
\EN 
where $m=N_{r'}-N_r$. 
The quotient of the set of finite transrations($\{^*\hat p_m(\SL_\RE)\}$) 
on $\SL$ by $^*\hat p_0(\SL_\RE)$, 
$$
\{^*\hat p_m(\SL_\RE)\}/^*\hat p_0(\SL_\RE),
$$ 
represents translations on $\SM$, which corresponds to translations on $\RE$. 

Since translations in higher dimensinal spaces are trivial, 
we do not discuss it here. 

\hfil\break
{\bf 5.2 Rotations}

Let us study rotations on two dimensional spaces $(\SL)^2$,
especially, rotations whose center 
is put at the origin of $(\SL)^2$, i.e., $\vec l_0(0)=(0,0)$. 
A rotation in two dimensional real space $\RE^2$, of which center is
at the origin, is represented by one parameter, i.e., a rotation 
angle $\T$. 
Under the rotation a point on $\RE^2$ written by $\vec r=(r{\rm cos}\A,
r{\rm sin}\A)$ is moved to $\vec r'=(r{\rm cos}(\A+\T),r{\rm sin}(\A+\T))$, 
where $r \in \RE$ and $0\leq \A, \T<2\pi$. 
The difference between the two vectors is 
\EQ
\vec r'-\vec r\equiv \vec d_\T(\vec r)=(d_\T^x(\vec r),d_\T^y(\vec r)),
\EN 
where 
\EQ
d_\T^x(\vec r)=r({\rm cos}(\A+\T)-{\rm cos}\A), \ \ \ 
d_\T^y(\vec r)=r({\rm sin}(\A+\T)-{\rm sin}\A).
\EN 
This means that the rotation of one point 
can be described by a displacement in 
the two dimensional space expressed  by 
$\vec d_\T(\vec r)$ such that 
\EQ
\hat R_\T \vec r\equiv \vec r'=\vec r+\vec d_\T(\vec r).
\EN 
We see that the rotation 
$\hat R_\T$ for the rotation angle $\T$ in $\RE^2$ can be described by a  
map from $\RE^2$ to $\RE^2$
producing the displacement $\vec d_\T(\vec r)$ for every vector $\vec r$. 

On two dimensinal infinitesimal-lattice subspace $(\SL_\RE)^2$, 
let us consider 
a map $^*\hat \RE_\T$
which transfers all vectors on $(\SL_\RE)^2$ to those on $(\SL_\RE)^2$ 
such that a vector 
  $\vec l_0(\vec r)= (l_0(r{\rm cos}\A),l_0(r{\rm sin}\A))$ is transferred  
to $\vec l_0(\vec r')=(l_0(r{\rm cos}(\A+\T)),l_0(r{\rm sin}(\A+\T)))$. 
($l_0(x)$ is defined in (17).) 
The displacement vector produced by the map 
is given by
\EQ
^*\hat R_\T \vec l_0(\vec r)-\vec l_0(\vec r)\equiv
\   ^* \vec d_\T(\vec r)= (\SDX,\SDY),
\EN 
where 
\EQ
\SDX=N_\T^x(\vec r) \SE, \ \ \ \SDY=N_\T^y(\vec r)\SE.
\EN
In (33) the integers $N_\T^x(\vec r),N_\T^y(\vec r)\in \ ^*{\cal Z}$ 
must fulfill the relations 
\EQN
N_\T^x(\vec r) \SE &\leq& r({\rm cos}(\A+\T)-{\rm cos}\A)<(N_\T^x(\vec r)+1) \SE, \nonumber\\
N_\T^y(\vec r) \SE &\leq& r({\rm sin}(\A+\T)-{\rm sin}\A)<(N_\T^y(\vec r)+1) \SE.
\ENN 
From one-to-one correspondence between ${\rm Mon}(r{\rm cos}\A|\SL)$ and 
${\rm Mon}(r{\rm cos}(\A+\T)|\SL)$ and that between 
${\rm Mon}(r{\rm sin}\A|\SL)$ and ${\rm Mon}(r{\rm sin}(\A+\T)|\SL)$ 
with respect to the above transfer (see the argument of $\S$5.1), 
one-to-one correspondence between  ${\rm Mon}(\vec r|\SL)$ and 
  ${\rm Mon}(\vec r'|\SL)$ is self-evident, 
where  ${\rm Mon}(\vec r|\SL)$ is defined by the set of
lattice-points $\ ^*\vec r=
(\ ^*x,\ ^*y)$ satisfying the relations $\ ^*x\in  {\rm Mon}(r{\rm cos}\A|\SL)$
and  $\ ^*y \in  {\rm Mon}(r{\rm sin}\A|\SL)$. 
Since the magnitude of $\vec r$ does not change 
under this map $^*\hat R_\T$, i.e., $|\vec r|=|\vec r'|$, 
all points on $(\SL_\RE)^2$ are mapped on $(\SL_\RE)^2$. 
We may consider that these maps $^*\hat R_\T$
for $\forall \T\in \RE$ ($0\leq \T<2\pi$) 
represent rotations for real angles on $(\SL_\RE)^2$, 
  which correspond to  the rotations $\hat R_\T$  on $\RE^2$. 

As was shown in $\S$5.1, $^*\DTX$ and $^*\DTY$ may be added by infinitesimal 
displacements as $k^{x}$$ \SE,\ k^{y}$$ \SE \in {\rm Mon}(0)$ for 
$k^x,\ k^y\in \ ^*{\cal Z}$. 
Note that these infinitesimal displacements do not depend on vecotrs $\vec r$. 
Maps producing these infinitesimal displacements 
represent no rotation on $(\SM)^2$ 
and we write the set of these maps by $^*\hat R_0(\SL_\RE)$.  
An arbitrary rotation $^*\hat R_\T(\SL_\RE)$  
on $(\SL_\RE)^2$ is represented by a map producing 
the displacement $^*\vec d_\T(\vec r)+\Delta \vec d_{\vec k}$ 
with $\Delta \vec d_{\vec k}=(k^x\SE,\ \ k^y\SE)$
for each vector $\vec l_n(\vec r)$. 
Actually  $\ ^*\hat R_\T(\SL_\RE)$ on $(\SL_\RE)^2$, 
which produce rotations for the fixed angle $\T\in\RE$ on $(\SM)^2$, 
are expressed by the sum
\EQ
^*\hat R_\T(\SL_\RE)=\ ^*\hat R_\T+\ ^*\hat R_0(\SL_\RE),
\EN 
where $^*\hat R_\T$ is defined in (32). 
Thus the rotations on $(\SM)^2$ are represented by 
\EQ
\{^*\hat R_\T(\SL_\RE)\}/\ ^*\hat R_0(\SL_\RE).
\EN 
where $\{^*\hat R_\T(\SL_\RE)\}$ stands for
the set of rotations $^*\hat R_\T (\SL_\RE)$  on $(\SL_\RE)^2$. 
It is obvious that these rotations make a group. 

The extension of these rotations to those
in higher dimensional spaces is trivial. 

\vskip10pt 
\hfil\break
{\large {\bf 6. Confined fractal-like property of $\SL$}}
\vskip10pt

We have shown that $\SM\cong \RE$. 
We, however, know that there is a large difference between them, that is, 
$\SM$ is constructed from the monad lattice-spaces $\MON$ which
contain infinite number of different lattice-points on $\SL_\RE$. 
In fact the power of $\MON$ can be not countable but continuous in general. 
Let us study the structure of $\MON$ in more details. 
We can write the elements of $\MON$ as 
\EQ
l_n(r)=(N_r+n)\SE,
\EN 
where, even if $N_r$ is fixed, $n$ can be elements of $\SN-\NA$, which 
satisfy the relation $n \SE\in {\rm Mon}(0)$. 
There are a lot of different possibilities depending on the choice of 
the original non-standard natural number $ ^*N \in\SN-\NA$. 
We shall here show two examples, that is, 
one has an infinite series of $\SM$ and the other a finite series.
(See Ref.11.) 
\vskip1pt
\hfil\break
{\bf 6.1 Infinite series of $\SM$}

Define an infinite series of infinite non-standard natural numbers
by the following ultra-products; 
\EQ
^*N_M\equiv \prod_{n\in\NA}\A_n^{(M)},\ \ \ {\rm for}\ \ M\in \NA 
\EN 
where $ \A_n^{(M)}=1$ for $0\leq n \leq M$ and $\A_n^{(M)}=(n+1)^{n-M}$ 
for $n>M$. 
Following the definition of the order $>$ for ultra-products, 
  we see that all of $^*N_M$ are infinity and the order 
is given by
\EQ
^*N_0>\ ^*N_1>\ ^*N_2>\cdots.
\EN 
Then we have an infinite series of infinitesimal numbers
\EQ
\SE_0< \ \SE_1<\ \SE_2<\ \cdots,
\EN 
where $\SE_M\equiv (^*N_M)^{-1}$. 
We can also prove that ratios 
\EQ
^*\lm_M\equiv {\ ^*N_{M-1} \over \ ^*N_M}, 
\ \ \ {\rm for}\ \ M\geq 1
\EN
are infinities of $\SN$. 
Since $^*N_0$ is an element of natural numbers 
$\SN-\NA$, we can take 
\EQ
\SE=\SE_0.
\EN 
Here let us consider the following rescaling for the lattice points;
\EQ
l_n(r)-l_0(r)=n\SE_0\equiv \ ^*\lm_1^{-1}l_n^{(1)}, 
\EN 
where 
\EQ
l_n^{(1)}=n\SE_1.
\EN 
Note that $l_n^{(1)}$ is independent of $r$. 
Even if the relation $n\SE_0\in {\rm Mon}(0)$ must be satisfied, 
the set of $n \in \SN$ satisfying the relation 
contains non-standard integers such that 
\EQ
n_m^{(1)}\equiv m\times \ ^*N_1 \in \ ^*{\cal Z}, 
\ \ \ {\rm for}\ \ \forall m\in {\cal Z}. 
\EN 
It is trivial that the relation is satisfied as 
\EQ
n_m^{(1)}\SE_0=m\lm_1^{-1}\in {\rm Mon}(0).
\EN 
It is also obvious that 
\EQ
n_m^{(1)}\SE_1=m\ \in {\cal Z}.
\EN 
Thus we can see that the set of $\forall l_n^{(1)},\ \ \SL_\RE^{(1)}
\equiv \{l_n^{(1)}=n\SE_1|n\in \ ^*{\cal Z},\ n\SE_0\in {\rm Mon}(0)\}$, 
is an infinitesimal-lattice space with the lattice-length $\SE_1$. 
In fact the set $\SL_\RE^{(1)}$ is constructed from
the elements of $\MON$ rescaled by the factor $\ ^*\lm_1$.  
From the facts that $\SL_\RE^{(1)}$ contains all integers, Archimedian 
property certifies the existence of natural numbers $m \geq |r|$ for $\forall r
\in \RE$ and the distance between two neighboring lattice-points is 
an infinitesimal number $\SE_1$, we can find 
an element of $\SN-\NA$, $N_r^{(1)}$, satisfying the relation 
\EQ
N_r^{(1)}\SE_1\leq r^{(1)}<(N_r^{(1)}+1)\SE_1,\ \ \ {\rm for}\ \ 
\forall r^{(1)}\in\RE.
\EN 
Following the same argument for the construction of $\SM$ given in $\S$4, 
we can introduce the monad of $r^{(1)}$, Mon($r^{(1)}|\SL_\RE^{(1)}$), 
by the set of the following lattice-points on $\SL_\RE^{(1)}$;
\EQ
l_n^{(1)}(r^{(1)})=(N_r^{(1)}+n^{(1)})\SE_1,
\EN 
where $n^{(1)}\in \ ^*{\cal Z}$ and st$(n^{(1)}\SE_1)=0$ must be fulfilled. 
It is obvious that Mon($r^{(1)}|\SL_\RE^{(1)}$) contains an infinite number of 
elements. 
Now we can define $\SM^{(1)}$ by the set 
\EQ
\SM^{(1)}\equiv \SL_\RE^{(1)}/{\rm Mon}(0|\SL_\RE^{(1)}). 
\EN 
The relation 
\EQ
\SM^{(1)}\cong \SM\cong \RE
\EN 
as additive groups is obvious. 
Thus translations and rotations on $N$-dimensional space $(\SM^{(1)})^N$ 
are described as same as those of $(\SM)^N$. 
We can conclude that every monad lattice-space 
$\MON$ for $\forall r\in \RE$ contain the 
same space $\SM^{(1)}$ by means of the same scale transformation. 

The second rescaling by using $^*N_2$ is carried out by following the 
same procedure presented above. 
We can perform the second rescaling by 
\EQ
l_n^{(2)}\equiv \ ^*\lm_2 (l_n^{(1)}(r^{(1)})-l_0^{(1)}(r^{(1)})),
\EN 
The derivations of $\SL_\RE^{(2)}$ and 
$\SM^{(2)}\equiv \SL_\RE^{(2)}/{\rm Mon}(0|\SL_\RE^{(2)})$ 
are same as those given in the previous argument, and then we have 
\EQ
\SM^{(2)}\cong \SM^{(1)}\cong \SM\cong \RE.
\EN 
By using the infinite series of $^*N_M$ we can proceed the same argument for the
construction of $\SM^{(M)}$ and thus we obtain the infinite series of 
sets isomorphic to $\RE$ as additive group such that 
\EQ
\RE\cong \SM \cong \SM^{(1)} \cong \cdots \cong \SM^{(M)}\cong \cdots.
\EN 

\hfil\break
{\bf 6.2 Finite series of $\SM$}

We definite a finite series of infinite numbers
\EQ
^*N_l^L\equiv \prod_{n\in \NA} (n+1)^{L-l}, \ \ \ {\rm for}\ l=0,1,2,\cdots,L-1
\EN 
where $^*N_l^L\in \SN-\NA$. 
We also see that \EQ
^*\lm_l\equiv {\ ^*N_{l-1}^L \over \ ^*N_l^L}=\prod_{n\in \NA}(n+1) \in \SN-\NA.
\EN 
Following the same argument as that of the infinite series, we can construct 
a finite series of sets isomorphic to $\RE$ as additive group
\EQ
\RE\cong \SM\cong \SM^{(1)}\cong \cdots \cong \SM^{(L-1)}. 
\EN 
\vskip5pt

We have many different examples for deriving such series. 
Note that $\MON$ does not have the structure discussed above, 
if $\ ^*N$ defined by (7) is taken, that is, the case for $L=1$ in 
the above argument. 
\vskip5pt

From the above arguments we understand that the set of finite lattice-points 
on $\SL$, i.e., $\SL_\RE$ contains series of spaces $\SM^{(M)}
\cong \RE$, 
which are constructed by means of relevant series of rescalings. 
Thus we may say that $\SL_\RE$ has a property similar to so-called 
fractal property. 
Note that the scaling parameter $^*\lm_n$ in the rescaling 
\EQ
l_m^{(n)}(r)-l_{m'}^{(n)}(r)=\ ^*\lm_n(l_m^{(n-1)}(r)-l_{m'}^{(n-1)}(r)),
\EN 
is infinity. 
So the similarity between fractal property and the structure of $\SL_\RE$ 
cannot directly define on $\RE$. 
We may consider that the infinitesimal fractal-like property of $\SL_\RE$ 
cannot directly be  observed on $\SM$. 
This means that the infinitesimal fractal-like property is confined 
on $\SM$. 
Note here that $\SL$ itself contains infinite number of the same sets 
as $\SL_\RE$. 
Then we may say that $\SL$ itself has the infinitesimal fractal-like property. 

\vskip30pt
\hfil\break
{\large {\bf 7. Construction of fields on $\SM$}}
\vskip10pt

Let us construct fields on $ ^*{\cal M}$. 
In the construction of field theory on $ ^*{\cal M}$ 
we follow the next two fundamental principles:[12]
\hfil\break
(I) All definitions and evaluations should be carried out 
on the original space $ ^*{\cal L}$. 
\hfil\break
(II) In definitions of any kinds of physical quantities on $\ ^*{\cal M}$, 
all the fields contained in the same monad lattice-space Mon$(r| ^*{\cal L})$ 
should be treated equivalently. (Principle of physical equivalece) 
\hfil\break 
It should be noted that the principle (I) means that theories which we will 
make on $\SL$ is generally not the same as any extensions of standard 
theories 
which have been constructed on $\RE$. 
The principle (I) also tells us 
that all physical expectation values on $\RE$ are obtained 
by taking standard part maps (maps from $ ^*{\cal R}$ to ${\cal R}$)[1] of 
results calculated on $ ^*{\cal L}$. 
The principle (II) is considered as the realization of 
the equivalence for 
indistinguishable quantities in quantum mechanics on non-standard space.[3] 
This principle, principle of physical equivalence, determines projections 
of physical systems defined on $\SL$ to those defined on $\SM$. 
Taking account of the fact that all points contained in 
the same monad lattice-space Mon$(r| ^*{\cal L})$ 
cannot be experimentally distinguished, the equivalent treatment with 
respect to all quantities defined on these indistinguishable 
 points is a natural requirement in the construction 
of theories on $ ^*{\cal M}$. 

\hfil\break
{\bf 7.1 Fields on $\SL$}

Let us define two fields $A(m)$ and $\bar A(m)$ on every lattice point 
$r(m)$ on $\ ^*{\cal L}$, which follow the commutation relations 
\begin{equation}
[A(m),\bar A(m')]=\delta_{mm'}\ \ \ {\rm and\ \  others}=0.
\end{equation}
The vacuum $| ^*0>=\prod_m|0>_m$ and the dual vacuum ${ <\ ^*\bar 0|}
=\prod_m\ _m<\bar 0|$ 
are defined by 
\begin{equation}
A(m)|0>_m=0\ \ \ \ {\rm and} \ \ \ \  _m<\bar 0|\bar A(m)=0
\end{equation}
with $_m<\bar 0|0>_m=1$. 
The fields $A(m)$ and $\bar A(m)$ operate only on the vacuum $|0>_m$ and the 
dual vacuum $_m<\bar 0|$. 
Following the principle (I), 
all expectation values are imposed to be calculated 
on $ ^*{\cal L}$ such that 
$$
<\ ^*\bar 0|\hat {\cal O}(\{ A\}, \{\bar A\})| ^*0>\ \in \SR, 
$$
where $\hat {\cal O}$ is operator constructed from the sets of the fields 
$A(m)$ and $\bar A(m)$. 
Physical values are obtained by the standard part map as
$$
{\rm st}(<\ ^*\bar 0| \hat {\cal O} | ^*0>) \in    {\cal R}.
$$

\hfil\break
{\bf 7.2 Fields on $\SM$}

Following principle of physical equivalence (principle (II)), 
we define fields at every point on $ ^*{\cal M}$ 
as the following equivalent sum over all fields contained 
in Mon$(r| ^*{\cal L})$;
\EQN 
\varphi([r])&\equiv&\ ^*\sum_{l} A(N_r+l)/\sqrt{^*\sum_{l}1}, \ \ \ \ \ 
\nonumber\\
\bar  \varphi ([r])&\equiv&\ ^*\sum_{l} \bar A(N_r+l)/
\sqrt{^*\sum_{l}1},
\ENN 
where $^*\sum_l\equiv \sum_{l, ^*\varepsilon l\in {\rm Mon}(0)} $ and 
hereafter 
$[r]$ in $\varphi([r])$ always means the fact that the equivalent sum over 
Mon$(r| ^*{\cal L})$ 
expressed by $^*\sum_l$ is carried out in the definition of $\varphi ([r])$. 
Here the equivalent sum is just the expression of principle of physical 
equivalence. 
We can easily evaluate the commutation relation 
\EQN
[\varphi ([r]),\bar \varphi ([r'])]=\ ^*\delta_{rr'}&=&1 \ ({\rm for}\ \ r'=r),
  \nonumber\\
             &=&0  \ ({\rm for}\ \ r'\not=r).
\ENN 
Note that $r,r'\in {\cal R}$ but $ ^*\delta_{rr'}$ is not equal to the usual 
Dirac delta function $\delta(r-r')$. 
Complex fields on $ ^*{\cal M}$, which are represented by linear 
combinations certifying the same weight 
for all fields contained in Mon$(r| ^*{\cal L})$, are generally written by 
\EQN 
\varphi ([r];k)&=&\ ^*\sum_{l} e^{i\theta_l^k(r)} A(N_r+l)/\sqrt{^*\sum_{l}1},\ \ \ \ \nonumber\\
\bar \varphi ([r];k)&=&\ ^*\sum_{ l}
e^{-i\theta_l^k(r)} \bar A(N_r+l)/\sqrt{^*\sum_{l}1},
\ENN 
where
$$
\theta_l^k(r)=\theta_k(r)+2\pi l k/\ ^*\sum_{l}1
$$
with the constraint  $ ^*\varepsilon k\in {\rm Mon}(0)$ for 
non-standard integers $k$. 
They satisfy the commutation relations 
$$
[\varphi ([r];k),\bar \varphi ([r'];k')]=\ ^*\delta_{rr'}\delta_{kk'}
$$ 
${\rm and \ \ others}=0$. 
These fields are the Fourier components for the fields on Mon$(r| ^*{\cal L})$ 
and their component number is same as that of $A(N_r+l)$ and $\bar A(N_r+l)$ 
included in Mon$(r| ^*{\cal L})$, because the constraint for $k$, that is, 
$ ^*\varepsilon k\in {\rm Mon}(0)$, is same as that for $l$. 
Note that $\varphi ([r])$ and $\bar \varphi ([r])$ given in (61) 
correspond to the above fields with $k=0$ and $\theta_0(r)=0$. 
Experimentally the differences of the wave numbers $k$ are not observable, 
because their wave lengths are infinitesimal. 
It is stressed that fields on one point of $\SM$ have infinite degrees of 
freedom. 
General fields on $\SM$ are described by functions of these fields 
such that
\begin{equation}
\phi([r])=f(\{\varphi ([r];k)\}, \{\bar \varphi ([r];k)\}). 
\end{equation}

\hfil\break
{\bf 7.3 Extension to $N$-dimensional space} 

Extension of the above consideration to $N$-dimensional space $(^*{\cal M})^N$ 
is trivial. 
Note that one should not confuse the $N$ for the $N$-dimensions of the space 
with the $N_{r}$ for the lattice number corresponding to $r$ of $\RE$ 
(see (16)) in the following discussions. 
Every point of $( ^*{\cal L})^N$ is represented by a $N$-dimensional 
vector 
$$
\vec r^N(\vec m)\equiv(r_1(m_1),.....,r_N(m_N)), 
$$ 
where $r_i(m_i)=\ ^*\varepsilon(N_{r_i}+l_i)$ with st$( ^*\varepsilon N_{r_i})=r_i \in 
{\cal R}$ and $ ^*\varepsilon l_i \in {\rm Mon}(0)$ for $i=1,.....,N$. 
Fields with $N$-components at $\vec r^N(\vec m)$, 
$A_j(\vec m)$ and $\bar A_k(\vec m)$ $(j,k=1,....,N)$, 
are defined by the following commutation relations;
\begin{equation}
[A_j(\vec m), \bar A_k(\vec m')]=\delta_{jk}\prod_{i=1}^N\delta_{m_im'_i}
\end{equation}
  ${\rm for}\ \ j,k=1,....,N\ \ {\rm and \ \ others}=0$. 
We may consider that these $N$ number of fields describe the $N$ 
oscillators of a lattice point corresponding to $N$ different directions of 
the space. 
The fields on $( ^*{\cal M})^N$ are described as follows;
\begin{equation}
\varphi_j ([\vec r^N
U(\{\alpha(\vec r^N];\vec k^N)=\ ^*\sum_{l_1}\cdot\cdot\cdot
\ ^*\sum_{l_N} e^{i\sum_{s=1}^N\theta_{l(s)}^{k(s)}
(\vec r^N)} A_j(N_{r_1}+l_1,\cdot\cdot\cdot,N_{r_N}+l_N)/( ^*\sum_l 1)^{N/2}
\end{equation} 
and similar to $\bar \varphi _j([\vec r^N];\vec k)$. 
We again have the commutation relations
\EQ
[\varphi _j([\vec r^N];\vec k^N), \bar \varphi _l([\vec r'^N];\vec k'^N)]=
\delta_{jl}\prod_{i=1}^N( ^*\delta_{r_ir'_i}\ \delta_{k_ik'_i})
\EN 
${\rm and\ \  others}=0$.

\vskip10pt
\hfil\break
{\large {\bf 8. Internal symmetries on $(\SM)^N$ induced from 
the confined 

substructure $(\MON)^N$}}
\vskip10pt

Symmetries on $( ^*{\cal M})^N$ which is 
induced from the internal substructure 
$({\rm Mon}(r| ^*{\cal L}))^N$ are expressed by 
transformations $U_T$ which keep all expectation values unchanged such that
$$
<\ ^*\bar 0|\hat {\cal O}(\{A\},\{\bar A\})| ^*0>
=<\ ^*\bar 0|U_T^{-1} U_T\hat {\cal O}(\{A\},\{\bar A\})U_T^{-1} U_T| ^*0>. 
$$ 
In general the transformation $U_T$ will be represented by maps 
of fields $A_j(\vec m)\ (\bar A_j(\vec m))$ 
to a linear combination of the fields $A_k(\vec m)\ (\bar A_k(\vec m))
\ (k=1,\cdots,N)$ on $\SL$. 
If the operators $U_T$ do not change the structure of $(\SM)^N$, 
they can represent symmetries on $(\MON)^N$.  

\hfil\break
{\bf 8.1 Transformation opertors on internal subspaces $(\MON)^N$}

Let us start from the construction of 
transformation operators on an internal subspace contained in 
a point on $( ^*{\cal M})^N$ corresponding to a point 
$\vec r^N=(r_1,.....,r_N)$ on $\RE^N$. 
The transformations map 
fields $A_j(\vec r^N(\vec m))$ 
($\bar A_j(\vec r^N(\vec m))$) 
on every lattice-point 
($\vec r^N(\vec m)=(N_{r_1}+l_1,...,N_{r_N}+l_N)$) 
to linear combinations of fields 
$A_k(N_{r_1}+l_1',...,N_{r_N}+l_N')$ 
($\bar A_k(N_{r_1}+l_1',...,N_{r_N}+l_N')$) ($k=1,...,N$) 
on the lattice-points of the same subspace. 
Following principle of physical equivalence (principle (II)), we construct 
the following $N^2$-number of operators $\hat T_{jk}([\vec r^N])$ 
on $(\SM)^N$, which are again defined by the 
equivalent sum over all fields contained in the $N$-dimensional subspace 
$({\rm Mon}(r| ^*{\cal L}))^N$ as 
\begin{equation}
\hat T_{jk}([\vec r^N])
=\ ^*\sum_{l_1}\cdot\cdot\cdot\ ^*\sum_{l_N} 
\bar A_j(N_{r_1}+l_1,...,N_{r_N}+l_N)
                                A_k(N_{r_1}+l_1,...,N_{r_N}+l_N).
\end{equation} 
We easily obtain commutation relations 
$$ 
[\hat T_{jk}([\vec r^N]),A_l(\vec r'^N(\vec m))]=
-(\prod_{i=1}^N\ ^* \delta_{r_ir'_i})\delta_{jl}A_k(\vec r^N(\vec m)), 
$$ 
$$ 
[\hat T_{jk}([\vec r^N]),\bar A_l(\vec r'^N(\vec m))]=
(\prod_{i=1}^N\ ^* \delta_{r_ir'_i})\delta_{kl}\bar A_j(\vec r^N(\vec m)), 
$$ 
\EQ
[\hat T_{jk}([\vec r^N]),\hat T_{lm}([\vec r'^N])]
=(\prod_{i=1}^N\ ^* \delta_{r_ir'_i}) (\delta_{kl}\hat T_{jm}([\vec r^N])
                                  -\delta_{jm}\hat T_{lk}([\vec r^N])).
\EN
These operators $\hat T_{jk}$ can be recomposed into the following generators; 
\hfil\break
(1) $U(1)$-generator:
\begin{equation}
\hat J_0=\sum_{j=1}^N \hat T_{jj}.
\end{equation}
(2) $SU(N)$-generators:
\begin{equation}
\hat J_L=\sum_{j=1}^{L+1}g_j \hat T_{jj},\ \  \ \ {\rm for}\ \ L=1,...,N-1
\end{equation}
with the traceless condition $\sum_{j=1}^{L+1}g_j=0$ and 
\begin{equation}
\hat J_{jk}^{(1)}=\hat T_{jk}+ \hat T_{kj},\ \ \ \ 
\hat J_{jk}^{(2)}={1 \over i}(\hat T_{jk}- \hat T_{kj}), 
\end{equation}
$ {\rm for}\ \ j\not=k$. 
\hfil\break
For instance, we can represent them by well-known matrices including Pauli spin 
matrices ${\vec \sigma}$ for $N=2$ case as
$$
\hat J_0\Rightarrow {\bf 1},\ \  \hat J_1\Rightarrow \sigma_3,\ \  \hat J_{12}^{(1)}
\Rightarrow \sigma_1 \ \ {\rm and}\ \ \hat J_{12}^{(2)}\Rightarrow \sigma_2.
$$ 

Now it is trivial that operators given by 
\begin{equation})\})={\rm exp}[i\sum_{j=1}^N\sum_{k=1}^N \alpha_{jk}(\vec r^N)\hat T_{jk}
([\vec r^N])]
\end{equation}
with st(
$
\forall \alpha_{jk}(\vec r^N))\in\ {\cal C}
$ 
(the set of complex numbers) produce maps of all fields on the subspace 
$(\MON)^N$ to linear combinations of the fields 
on the same subspace. 
From the construction procedure of $\hat T_{jk}$ 
it is obvious that the operators do not break the structure of $(\SM)^N$. 
Note also that $U$ does not change the vacuum and the dual vacuum, because 
$$
\forall \hat T_{jk}| ^*0>=< ^*\bar 0|\forall \hat T_{jk}=0.
$$

\hfil\break
{\bf 8.2 Symmetries on $(\SM)^N$}

Operators on $(\SM)^N$ can be defined by products of $U({\A(\vec r^N)})$ 
as
\EQ
U_T(\{\A\})=\prod_{i=1}^N \ ^*\prod_{N_{ri}}U(\{\A(\vec r^N)\}), 
\EN 
where $\ ^*\prod_{N_{ri}}$ stand for the product with respect to 
$\forall N_{r_i}$ 
with the constraint st$(\SE N_{r_i})=r_i\in\RE$ . 
(See the definition of $\SE N_{r_i}$ given in (16).) 
It is interesting that the transformations produced by $U_T(\{\alpha\})$ 
are generally local transformations on our observed space $( ^*{\cal M})^N$ 
because the parameters $\{\alpha\}$ can depend on the position $\vec r^N$,   
whereas they are global ones on the internal subspace 
$({\rm Mon}(r| ^*{\cal L}))^N$. 
Note that $U_T$ does not change the vacuum and the dual vacuum.

Let us show a few realistic transformations included in $U_T$. 
\hfil\break
(a) $U(1)$ transformation:
\begin{equation}
U_0(\vec r^N)={\rm exp}[i\alpha_0(\vec r^N)\hat J_0([\vec r^N])]
\end{equation} 
${\rm for\ \  st}(\alpha_0)\in  {\cal R}$. 
It is an interesting problem to investigate whether  this $U(1)$ symmetry can be 
the $U(1)$ symmetry of electro-weak gauge theory or the solution of so-called 
$U(1)$ problem in hadron dynamics. 
\hfil\break 
(b) $SU(N)$ transformation:
\begin{equation}
U_N(\vec r^N)={\rm exp}[i\{\sum_{L=1}^{N-1}\alpha_L(\vec r^N)\hat J_L([\vec r^N])
                +\sum_{j=1}^{k-1}\sum_{k=2}^{N}\sum_{i=1}^2\alpha_{jk}^{(i)}(\vec r^N)
                  \hat J_{jk}^{(i)}([\vec r^N])\}]
\end{equation} 
for st$(\forall \alpha_L), \ {\rm st }( \forall \alpha_{jk}^{(i)}) \in {\cal R}$. 
It is an interesting proposal 
that three color components of QCD may be identified by those of $U_3(\vec r^3)$
for three spatial dimensions.

\vskip10pt
\hfil\break
{\large {\bf 9. Quantized configuration space and infinitesimal distances}} 
\vskip10pt

Here we study configuration space describing 
$ ^*{\cal M}$, which will be 
useful in the investigations of general relativity and gravitaions. 

\hfil\break
{ \bf 9.1 Quantization of configuration space}

Let us start from 1-dimensional space. 
We can construct  position operator 
\EQ 
\hat r_{\SM}=\ ^*\sum_{N_r} r \hat T_r,
\EN
where $\ ^*\sum_{N_r}$ stands for the sum over $\forall N_r$ with the 
constraint st$(\SE N_r)=r\in\RE$ and 
\EQ 
\hat T_r=\ ^*\sum_l \bar A(N_r+l)A(N_r+l).
\EN 
Following 
principle of physical equivqlence, 
$\hat T_r$ is expressed by the equivalent sum with respect to 
all fields in the same monad lattice-space $\MON$.  
Note that $r$ in (77) can be replaced by $r+a_r \SE$ with the constant 
st$(a_r \SE)=0$ for $\forall r\in \RE$. 
The eigenstate of $\hat r_{\SM}$ for the eigenvalue $r$ is written by 
\EQ 
|r>_{\SM}\equiv  \bar \varphi([r])
| ^*0>. 
\EN 
Hereafter we call them monad states. 
The relation
$$
\hat r_{\SM}|r>_{\SM}=r|r>_{\SM}
$$ 
is trivial. 
If one does not want to have 0 eigenvalue for $r=0$, $r+a_r \SE$ can be used 
instead of $r$ in the definitoin of $\hat r_{\SM}$. 
The monad states $|r>_{\SM}$ are quite similar to the ket states 
of usual quantum mechanics except the normalization condition 
$$
\ _{\SM}<r|r'>_{\SM}=\ ^*\delta_{rr'}, 
$$ 
where $_{\SM}<r|=<\ ^*\bar 0|\ ^*\prod_{N_r}\varphi([r])$. 
It is noted that every monad state $|r>_{\SM}$ has its own 
internal substructure $\MON$. 
\vskip5pt

Now we can define the quantized states for our configuration space as follows;
\EQ 
|\SM>\equiv \ ^*\prod_{N_r} |r>_{\SM},\ \ \ <\SM|\equiv \ ^*\prod_{N_r}\ _{\SM}<r|.
\EN 
On these states the position operator $\hat r_{\SM}$ is represented 
by a diagonal operator 
and then we can consider that the base state $|\SM>$ describes our configuration space, which is normalized as 
$<\SM|\SM>=1$. 

Extension to $N$-dimension is trivial. 
A component of the position-vector operator can be defined as same as 
that of the 1-dimensinal case, e.g., for the $i$th component 
\EQ 
\hat r_{i\SM}=\ ^*\sum_{N_{r1}}\cdots\ ^*\sum_{N_{rN}} r_i 
\hat T_i([\vec r^N]),
\EN 
where 
\EQ 
\hat T_i([\vec r^N]) =\ ^*\sum_{l_1}\cdots\ ^*\sum_{l_N} 
\bar A_i(N_{r_1}+l_1,...,N_{r_N}+l_N)A_i(N_{r_1}+l_1,...,N_{r_N}+l_N) 
\EN 
for $i=1,2,...,N$. 
The $N$-dimensional configuration state is expressed by 
\EQ 
|\SM^N>=\prod_{j=1}^N(\ ^*\prod_{N_{r1}}\cdots\ ^*\prod_{N_{rN}}
 \bar \varphi_j([\vec r^N]))| ^*0>. 
\EN


\hfil\break
{\bf 9.2 Infinitesimal distance}

Infinitesimal relative distance operators are definable only on the internal 
subspace Mon$(r| ^*{\cal L})$ such that
\begin{equation}
d\hat r(\Delta l)\equiv \hat r(N_r+l)-\hat r(N_r+l'),
\end{equation} 
where 
$ \Delta l\equiv l- l'$ and 
$$ 
\hat r(N_r+k)\equiv\ ^*\varepsilon (N_r+l)  \bar A([r]) A(N_r+k)
$$ 
with the definiion 
$$ 
\bar A([r])\equiv \ ^*\sum_l \bar A(N_r+l),
$$ 
which follows princilpe of physical equivalnce. 
The monad states 
$|r>_{\SM}$ are the eigenstates of $\hat r(N_r+l)$ and $d\hat r(\Delta l)$. 
We actually obtain 
\begin{equation}
d\hat r(\Delta l)|r>_{\SM}=\ ^*\varepsilon \Delta l|r>_{\SM}. 
\end{equation} 
\vskip5pt

We can write squared distance operators in the $N$-dimensional space  as 
\begin{equation}
(d \hat s)^2(\vec r^N)=d \hat r_\mu (\Delta \vec l^N) g^{\mu\nu}d \hat r_\nu
(\Delta \vec l^N),
\end{equation} 
where the sums over $\mu$ and $\nu$ from $1$ to $N$ are neglected, 
$$
d \hat r_\mu (\Delta \vec l^N)=
\hat r_\mu(N_{r_1}+l_1,...,N_{r_N}+l_N)-
\hat r_\mu(N_{r_1}+l'_1,...,N_{r_N}+l'_N)
$$ with 
$\hat r_\mu(N_{r_1}+l_1,...,N_{r_N}+l_N)=\ ^*\varepsilon (N_{r_\mu}+l_\mu)
\bar A_\mu([\vec r^N])A_\mu(N_{r_1}+l_1,...,N_{r_N}+l_N)$ 
and 
$
\Delta \vec l^N=(l_1-l'_1,\cdot\cdot\cdot,l_N-l'_N). 
$
If the metric oprator $g^{\mu\nu}$ is taken as Minkowski metric,
the internal subspace $({\rm Mon}(r| ^*{\cal L}))^N$ 
just represents so-called local inertial system in general relativity. 
We have the equations 
\EQN 
d \hat r_\mu (\Delta \vec l^N)|\vec r^N>_{\SM}&=&
\SE\Delta l_\mu |\vec r^N>_{\SM}, \CR
(d\vec s)^2(\vec r^N)|\vec r^N>&=&\SE^2 \Delta l_\mu g^{\mu\nu} \Delta l_\nu
|\vec r^N>_{\SM}. 
\ENN 
The expectation value of $(d \hat s)^2$ is calculated as follows;
$$
(ds)^2=_{\SM}<\vec r^N|(d \hat s)^2(\vec r^N)|\vec r^N>_{\SM}.
$$ 
The same expectaton value of the squared distance operator can be 
obtained in terms of the expectation value 
with respect to the configuration state 
$|\SM^N>$. 
It is transparent 
that transformations keeping $(ds)^2$ unchanged 
are represented by $U(\{\alpha(\vec r^N)\})$ given in (73).

\vskip10pt
\hfil\break
{\large {\bf 10. Translations, Rotations and 
Lorentz and general relativistic transformations}}
\vskip10pt

Let us study symmetries on the configuration space, 
which keep all expectation values  
unchanged such that 
$$
<\SM^N|U^{-1}U \hat {\cal O}(\{\bar A\},\{A\})U^{-1}U|\SM^N>.
$$ 
Note that the configuration state $|\SM^N>$, the dual state $<\SM^N|$ 
and operators 
are transformed as follows; 
$$
|\SM^N>\longrightarrow U|\SM^N>,\ \ \ 
<\SM^N|\longrightarrow <\SM^N|U^{-1}\ \ \ , 
U \hat {\cal O}(....)U^{-1}.
$$ 

\vskip5pt
\hfil\break
{\bf 10.1 Translational invariance on $(\SM)^N$}

The operator which replaces $|r>$ with $|r+\Delta>$ for $\Delta \in \RE$ 
is obtained as 
\EQ 
\hat p_r(\Delta)=\ ^*\sum_l \bar A(N_{r+\Delta}+l)A(N_r+l). 
\EN 
We have $\hat p_r(\Delta)| ^*0>=0$. 
Then we can define the translation operator  by 
\EQ 
\hat P(\Delta)= :\ ^*\prod_{N_r} \hat p_r(\Delta):, 
\EN 
where $:......:$ means the normal product used in usual field theory, 
in which all creation operators ($\bar A_j(m)$) must put on the left-hand 
side of all annihilation operators ($A_j(m)$). 
We see that $\hat P(\Delta)$ transforms the configuration state $|\SM>$ 
to the isomorophic space, that is, 
\EQ 
\hat P(\Delta)|\SM>\cong |\SM>
\EN 
for $\forall \Delta \in \RE$. 

Let us study the invariance of expectation values 
$$
<\SM|\hat {\cal O}(\{\bar A\}, \{A\})|\SM>. 
$$ 
Taking account of the definitions of
$|\SM>=\prod \bar \varphi ([r])|\ ^*0>$ and 
$<\SM|=<\ ^*\bar 0|\prod \varphi([r])$ 
and the fact that 
all the fields commute each other except $A$ and $\bar A$ on the same 
lattice-point, the number of $A$ and that of $\bar A$ 
on the same lattice-point must be same in operators having non-vanishing 
expectation values on $|\SM>$. 
This means that  every term of such operators 
must be written by 
the product of powers 
such as $(\bar A A)^n$ with $n\in \NA$ for all pairs of $A$ and 
$\bar A$ on the same lattice-point. 
On the other hand we easily see that the products of $\bar A A$ on the same 
lattice-piont commute with $\hat P(\Delta)$ such that 
\EQ 
[\bar A A,\hat P(\Delta)]=0
\EN 
for $\forall \Delta \in \RE$. 
Now we can conclude that operators having non-vanishing expetation values 
commute with the translation operators, that is, 
\EQ 
[\hat {\cal O}(\{\bar A\}, \{A\}),\hat P(\Delta)]=0. 
\EN 
Translational invariance is certified for physically meaningful operators 
as 
\EQN 
<\SM|\hat P(-\Delta)\hat {\cal O}(...)\hat P(\Delta)|\SM>&=&
<\SM|\hat P(-\Delta) \hat P(\Delta) \hat {\cal O}(...)|\SM>   \CR
&=&<\SM|\hat {\cal O}(...)|\SM> 
\ENN 
because of the relation 
$<\SM|\hat P(-\Delta) \hat P(\Delta)=<\SM|$. 

The extension of the above argument to the $N$-dimensional spaces is 
trivial. 
Note also that the translations cannot be generated by the operators 
$U_T$ given in (74). 

\vskip5pt
\hfil\break
{\bf 10.2 Rotations}

Rotational invariance can be introduced only for subspaces whose metric 
$g^{\mu\nu}$ have the same sign like $SO(3)$ subspace of 
$SO(3,1)$. 
Generators for the rotations in $(j,k)$-plane are given by 
\EQ 
\hat J_{jk}=\hat T_{jk}-\hat T_{kj}. 
\EN 
In general rotation operators are described by 
\EQ 
U_R(\{\theta\})=e^{i\sum_{(j,k)}\theta_{jk} \hat J_{jk}}. 
\EN 
It is transparent that $U_R$ for st$(\forall \theta_{jk}) \in \RE$ 
are unitary operators and generate rotations 
on the subspace. 

\vskip5pt
\hfil\break
{\bf 10.3 Lorentz transfomations}

Position operator for one point on $(\SM)^N$ corresponding to 
$\vec r^N$ on $\RE^N$ is given by 
\EQ 
\hat r_j([\vec r^N])=r_j \bar \varphi_j([\vec r^N]) \varphi_j([\vec r^N]), 
\ \ {\rm for}\ j=1,...,N. 
\EN 
The expectation value of squared distance from the origin are 
evaluated as 
\EQ 
(\vec r^N)^2=<\SM^N|\hat r_\mu ([\vec r^N])g^{\mu\nu}\hat r_\nu([\vec r^N]) 
|\SM^N>, 
\EN 
where the metric tensors $g^{\mu\nu}$ are taken as Minkowski metric tensors. 

Let us study the simplest case for $N=2$. 
The metric tensors are chosen such that 
$$
g^{11}=-g^{22}=1 \ \ {\rm and}\ \ g^{12}=g^{21}=0.
$$ 
Transformations 
\EQ
U_L(a)=\prod_{j=1}^N \ ^*\prod_{N_{rj}} e^{-a\hat J_{12}^{(1)}([\vec r^N])}
\EN 
with the constraint st$(a) \in \RE$ 
(see (73) and (74)) generate 2-dimensional Lorentz 
transformations which are expressed in 2-dimensional matrices as 
\[   U_L(a)=
\left(
   \begin{array}{cc}
          \ \ {\rm cosh}a & \ \ -{\rm sinh}a   \\
           \ -{\rm sinh}a & \ \ \ \ {\rm cosh}a 
   \end{array}
\right)
\]

Generalization for the $N$-dimensions can be performed by using 
combinations of $U_L(a)$ with the rotations.

\vskip5pt
\hfil\break
{\bf 10.4 General relativistic transformations} 

We have many different types of transformations which 
keep the squred distance $(\vec r^N)^2$ invariant but generally do not 
the metric tensors invariant, while Lorentz transformations keep both 
of them invariant. 
They are described by the transformations $U_T(\{\A\})$ given in (74), where 
the parameters $\{\A\}$ should be chosen such that all the axes are real 
after the translations.  
Of course, all the parameters must be finite. 
In such transformations we have different types of vectors corresponding 
to covariant and contravariant tensors in general coordinate transformations. 
The difference between them is expressed 
as follows; 
\EQN 
U_G \hat r_\mu|\SM^N>, &\ \ &{\rm for\  covariant\ vectors}  \CR
<\SM^N|\hat r_\mu g^{\mu\nu}U_G^{-1}, &\ \ &{\rm for\ contravariant\ vectors}.
\ENN 

A simple example representing dilatation transformations 
are described by 
\EQ 
D_d=e^{-\sum_{j=1}^N a_j(\vec r^N)\hat T_{jj}([\vec rN])}, 
\EN 
which transforms as
$$
U_d\ \hat r_\mu|\SM^N>=e^{-a_{\mu}(\vec r^N)}\hat r_\mu|\SM^N>,
$$ 
$$
<\SM^N|\hat r_\nu g^{\nu\mu}U_d^{-1}=<\SM^N|\hat r_\nu g^{\nu\mu}
e^{a_\mu(\vec r^N)}. 
$$

\vskip5pt

Note again that $U_G(\{\alpha(\vec r^N)\})$ is global on the subspace 
$(\MON)^N$, 
even though it is generally local on observed space $(\SM)^N$. 
We understand that all the transformations described by $U_T$ 
of (74) can include general relativistic transformations. 
This fact implies that  general relativstivc transformations 
are generally represented by local non-abelian transformations.

\vskip10pt
\hfil\break
{\large {\bf 11. Remarks on fermionic oscillators}}
\vskip10pt

In this section we shall comment 
that instead of bosonic fields $A(m)$ and $\bar A(m)$ 
we can construct similar field theory 
by using fermionic fields $C(m)$ and $\bar C(m)$ which satisfy anticommutaion 
relations $[C(m),\bar C(m)]_+=1$ and commutation relations 
$[C(m),C(m')]_-=[C(m),\bar C(m')]_-
=[\bar C(m),\bar C(m')]_-=0$ for $m\not=m'$. 

As far as operators 
$\hat T_{jk}([\vec r^N])$ presented in (68) are 
concerned, we can define them by the replacement of $A$ and $\bar A$ with 
$C$ and $\bar C$, respectively. 
And we get the same commutation relations given in (69). 
This means that all the arguments of the internal 
symmetries performed in the bosonic oscillator case 
are completely accomplished in the fermionic oscillator case. 
That is to say, as far as the internal symmetries are concerned, 
there is no difference between the bosonic and the fermionic cases. 
Futhermore we can easily understand that not only $U_T$ but also all 
other operators written by the products 
of $\bar A$ and $A$ like $ \hat T_r,\ \hat r$ and $\hat p_r$ can be defined 
in the replacement of $\bar A\ A$ with $\bar C\ C$ and they have the same 
properties as discussed in the bosonic case. 

Difference between them appears in the construction of realistic fields 
from $\varphi([r];k)$. 
Namely products of more than the non-standard natural number 
$\ ^*\sum_l 1$ with respect to the fields 
$\varphi([r];k)$ vanish for the fermionic case, 
whereas there is no such restriction in the bosonic case. 
We may say that the concept of antiparticles will be introduced more easily 
in the fermionic case by using occupaton and unoccupation numbers of 
lattice-points of the monad lattice-space $\MON$. 

Anyhow the selection of the bosonic or the fermionic 
or both like supersymmetric 
is still open question at present.


\vskip10pt
\hfil\break
{\large {\bf 12. Concluding remarks}}
\vskip10pt

We have constructed a field thoery on the quantized space-time by using 
infinitesimal-lattice space $(\ ^*{\cal L})^N$. 
In this scheme the internal subspace $({\rm Mon}(r| ^*{\cal L}))^N$ and 
the symmety transformation $U_T$ 
induced from the subspace are uniquely determined, when we 
construct the field theory on $( ^*{\cal M})^N\cong {\cal R}^N$. 
Since all definitions and evaluations are imposed to be done on $(\ ^*{\cal L})^N$, 
we can perform them in terms of $*$-finte sum in non-standard analysis. 
In fact we need not introduce any Dirac $\delta$-functions. 
In this scheme we can carry out all evaluations on configuration space, 
not on Fock space in usual field theory. 
This fact is an interesting advantage in the investigation of quantum gravity, 
as was seen in the introduction of the infinitesimal relative distance and 
the local inertial system. 
In order to investigate this model in more detail an inevitable problem is 
introducing equation of motions on $( ^*{\cal M})^N$, 
which will be represented by difference equation on Mon$(r| ^*{\cal L})$. 
It is also interesting to study relations between the general field $\phi([r])$ 
and observed fields like leptons, quarks, gauge fields and etc. 
To carry out these works we have to investigate the symmetries described by $U_T$ 
more precisely. 
\vskip5pt

Finally I would like to present the global view of theory on non-standard 
space once more. 
The fundamental concept is introducing the equivalence based on 
experimental errors (physical equivalence) into theories 
in a mathematically consistent logic, which is allowed only on non-standard 
spaces. 
On the spaces the physical equivalence determine projections from 
non-standard spaces to observed spaces $\RE^N$, which are described by filters in non-standard theory. 
In fact the filters determine topologies, because they determine 
the structure of the monad space and then that of the observed space.[1] 
This means that we observe physical phenomena which depend on the errors, 
that is to say, we have to answer the following questions to 
determine the worlds which we observe in experiments: 

{\it Which quantities are taken as observables accompanied by errors 
in experiments?} 

{\it How large are the errors?} 
\hfil\break
We have to understand that in an experiment
 we are allowed to peep only through 
a filter which is determined by the physical 
equivalence based on the errors of the experiment.  
Theories on observed spaces, which explain experimental results, of course 
have to depend on 
the filters which determine the projections of the theory on the 
non-standard space to 
theories on the observed spaces, 
even if the theory is uniqe on the non-standard space. 
Actually we have presented some different filters, for instance,  

filters with $\hbar \in {\rm Mon}(0)$ derives classical limits,[6,7] 

filters with $\hbar \in {\rm Mon}(0)$ but $\hbar N \not\in {\rm Mon}(0)$ 
does macroscopic limit,[8,9]

filters with $g$(coupling of objects with heat baths) $\in {\rm Mon}(0)$ 
does microcanonical 

ensembles of statistical mechanics.[2,3,4,10] 
\hfil\break
We see that those filters derive different monad spaces and then 
different observed spaces (different theories). 
In any quantum mechanical systems experimental errors are mainly determined 
by the characters of measurement apparatus, even though the erroes 
are produced from the interactions between objects and apparatus. 
We have to determine the filters in analyzing the schemes of the apparatus 
which produce the errors in the experiments. 
Here I would like again to repeat that we cannot perform any expriments 
which are not accompanied by any errors. 
Therefore we have always to take account of phenomena hidden behind 
experimental errors, when we make theories in our observed spaces. 

\vskip5pt 
\hfil\break
{\large {\bf References}}
\hfil\break
[1]  A. Robinson, Non Standard Analysis (North-Holland, Amsterdam, 1970).
\hfil\break
$\ \ \ $M. Saito, Ultra-Products and Nonstandard Analysis (Tokyo-Tosyo, Tokyo, 
1976) (in Japanese).\hfil\break
$\ \ \ $S. Albeverio et al., Nonstandard Method in Stochastic Analysis and 
Mathematical Physics (Academic Press, New York, 1985).\hfil\break
[2] T. Kobayashi,  Nuovo Cim. 113B (1998) 1407.
\hfil\break
[3] T. Kobayashi, Proceedings of 5th Wigner Symposium, edited by 
P. Kasperkovitz and D. Grau (World Scientific, Singapoe, 1998) 518.
\hfil\break
[4] T. Kobayahi, Symmetries 
in Science X, edited by B.Gruber and M. Ramek (Plenum Press, New York and London, 1998) 153. 
\hfil\break
[5] M. O. Farrukh, J. Math. Phys. 16(1975) 177.
\hfil\break
[6] T. Kobayashi, Symmetries in Science VII, edited by B.Gruber and T. 
Otsuka, (Plenum Press, New York, 1994) 287.
\hfil\break
[7] T. Kobayashi, Nuovo Cim. 110B (1995) 61.
\hfil\break
[8] T. Kobayashi,  Nuovo Cim. 111B (1996) 227.
\hfil\break
[9] T. Kobayashi, Proceedings of the Fourth International 
Conference on Squeezed States and Uncertainty Relations, edited by A. Han, 
(NASA Conference Publication 3322, 1996) 301. 
\hfil\break
[10] T. Kobayashi, Phys. Lett. A, 207(1995) 320; 210(1996) 241; 222(1996) 26.
\hfil\break
[11] T. Kobayashi, {\it Translastions, Rotations and Confined Fractal 
Property on Infinitesimal-Lattice Spaces}, preprint of University of Tsukuba 
(1997). 
\hfil\break
[12] T. Kobatashi, Talk in  the XI International Conference on Problems of Quantum 
Field Theory, July, 1998, Dubna, Russia (to appear in the Proceedings). 
\hfil\break
\end{document}